\documentclass[11pt,preprint,flushrt]{aastex}
\usepackage{epsfig, graphicx}

\newcommand{\um}{12}
\newcommand{\cit}{1}
\newcommand{\lbl}{2}
\newcommand{\stockholm}{5}
\newcommand{\lpnhe}{6}
\newcommand{\yale}{9}
\newcommand{\lam}{8}
\newcommand{\upenn}{10}
\newcommand{\ucb}{11}
\newcommand{\stsci}{12}
\newcommand{\cppm}{7}
\newcommand{\iu}{13}
\newcommand{\aas}{16}
\newcommand{\uc}{17}
\newcommand{\cambridge}{18}
\newcommand{\cea}{19}
\newcommand{\ipnl}{20}
\newcommand{\slac}{4}
\newcommand{\fnal}{3}
\newcommand{\anu}{15}
\newcommand{\jpl}{14}

\begin{document}

\title{Probing Dark Energy via Weak Gravitational Lensing with
the SuperNova Acceleration Probe (SNAP) \\ 
{\it A White Paper to the Dark Energy Task Force}}

\author{J.~Albert\altaffilmark{\cit},
G.~Aldering\altaffilmark{\lbl}, S.~Allam\altaffilmark{\fnal},
W.~Althouse\altaffilmark{\slac},
R.~Amanullah\altaffilmark{\stockholm},
J.~Annis\altaffilmark{\fnal}, P.~Astier\altaffilmark{\lpnhe},
M.~Aumeunier\altaffilmark{\cppm,\lam},
S.~Bailey\altaffilmark{\lbl}, C.~Baltay\altaffilmark{\yale}
% ...
}

\email{saul@lbl.gov} \email{melevi@lbl.gov}

\altaffiltext{\cit}{California Institute of Technology}
\altaffiltext{\lbl}{Lawrence Berkeley National Laboratory}
\altaffiltext{\fnal}{Fermi National Accelerator Laboratory}
\altaffiltext{\slac} {Stanford Linear Accelerator Center}
\altaffiltext{\stockholm}{University of Stockholm}
\altaffiltext{\lpnhe}{LPNHE, CNRS-IN2P3, Paris, France}
\altaffiltext{\cppm}{CPPM, CNRS-IN2P3, Marseille, France}
\altaffiltext{\lam}{LAM, CNRS-INSU, Marseille, France}
\altaffiltext{\yale}{Yale University}
\altaffiltext{\upenn}{University of Pennsylvania}
\altaffiltext{\ucb}{University of California at Berkeley}
\altaffiltext{\um}{University of Michigan}
\altaffiltext{\stsci}{Space Telescope Science Institute}
\altaffiltext{\iu}{Indiana University} \altaffiltext{\jpl}{Jet
Propulsion Laboratory} \altaffiltext{\anu}{The Australian National
University} \altaffiltext{\aas}{American Astronomical Society}
\altaffiltext{\uc}{University of Chicago}
\altaffiltext{\cambridge}{Cambridge University}
\altaffiltext{\cea}{CEA, Saclay, France}
\altaffiltext{\ipnl}{IPNL, CNRS-IN2P3, Villeurbanne, France}

\section{Project Summary:} SNAP is a candidate for the Joint Dark
Energy Mission (JDEM) that seeks to place constraints on the dark
energy using two distinct methods.  The first, Type Ia SN, is
discussed in a separate white paper.  The second method is weak
gravitational lensing, which relies on the coherent distortions in
the shapes of background galaxies by foreground mass structures.
The excellent spatial resolution and photometric accuracy afforded
by a 2-meter space-based observatory are crucial for achieving the
high surface density of resolved galaxies, the tight control of
systematic errors in the telescope's Point Spread Function (PSF),
and the exquisite redshift accuracy and depth required by this
project. These are achieved by the elimination of atmospheric
distortion and much of the thermal and gravity loads on the
telescope.  The SN and WL methods for probing  dark energy are
highly complementary and the error contours from the two methods
are largely orthogonal.

The nominal SNAP weak lensing survey  covers  1000 square degrees
per year of operation in six optical and three near infrared
filters (NIR) spanning the range 350 nm to 1.7$\mu$m. This survey
will reach a depth of 26.6 AB magnitude in each of the nine
filters and allow for approximately 100 resolved galaxies per
square arcminute, $\sim3$ times that available from the best
ground-based surveys.  Photometric redshifts will be measured with
statistical accuracy that enables scientific applications for even
the faint, high redshift end of the sample.  Ongoing work aims to
meet the requirements on systematics in galaxy shape measurement,
photometric redshift biases, and theoretical predictions.


\section{Weak Lensing as a Probe of Dark Energy}\label{probe}

The study of modern cosmology has been tremendously advanced by
probes for which detailed comparison of theory and observation is
possible. Weak gravitational lensing (WL) is one such probe,
combining theoretical control and experimental tractability with
sensitivity to interesting cosmological parameters (see Refregier
2003 for a review). WL is the slight distortion in the images of
distant source galaxies due to deflections of light rays by the
inhomogeneous mass distribution of the Universe.  The size of
these distortions depends upon both $D(z)$, the distances
traveled, and upon $G(z)$, the growth function which determines
the amplitude of the deflecting mass concentrations.  WL is an
attractive cosmological probe because the physical effect,
gravitational deflection of light, is simple and well understood.
Furthermore, the deflecting masses are dominated by dark matter,
the evolution of which is purely gravitational and hence
calculable.

The sensitivity of WL to both distances and gravitational growth
makes lensing a particularly powerful probe of dark energy,
allowing us to differentiate models of dark energy from
modifications to Einstein gravity (Linder 2005). WL power spectra
and cross-spectra among 5-10 bins in redshift are the most robust
measure of dark energy. Recent work has shown that the weak
lensing sky contains information beyond the power spectrum:
non-Gaussian signatures captured through the bispectrum (BS;
3-point shear correlations), highly non-linear features (galaxy
clusters), and cross-correlations with foreground galaxies. These
three additional probes of dark energy with WL could be extremely
powerful in conjunction, but their theory and measurement needs
further investigation.

Because the weak lensing induced distortions on the shapes of
individual galaxies are small ($\sim 0.1-2\%$) compared to the
intrinsic scatter in the shapes of the galaxies ($\sim 30\%$),
weak lensing must be measured statistically. A weak lensing
measurement needs to overcome both this intrinsic scatter in
galaxy shapes (by imaging a high surface density of resolved
galaxies), and the noise on the measurement of each galaxy shape.
This noise is predominantly due to the PSF (of the atmosphere,
telescope, etc.).  Lack of systematics control or a reduced number
density of accurately resolved galaxies will limit the statistical
impact of increased area.

To extract cosmological information from WL we need to:
\begin{itemize}
\item Measure the shapes and photometric redshifts (photo-z's)
of very large numbers ($10^8$ or more) of
background galaxies on the sky.
\item Remove the effect of telescope or atmospheric optics on the
shapes of these galaxies.
\item Develop statistics that extract cosmological information from a
catalog of galaxy shapes and redshifts.
\end{itemize}
The advantage of space-based weak lensing dark energy measurements
comes from the extent to which systematic errors in shape and
redshift measurement can be controlled; ground-based observations
have far larger and more variable optical distortions than would
SNAP. Further, as shown below, poorer angular resolution makes
ground-based surveys more susceptible to small errors in PSF
treatment. Space provides important gains with the  higher number
density of galaxies, better photometric redshifts, and a broader
redshift coverage which enables studies at smaller scales and
higher redshift than possible from the ground.

\section{The SNAP Weak Lensing Survey}

\label{survey} 
\begin{figure}[h]
 \plotone{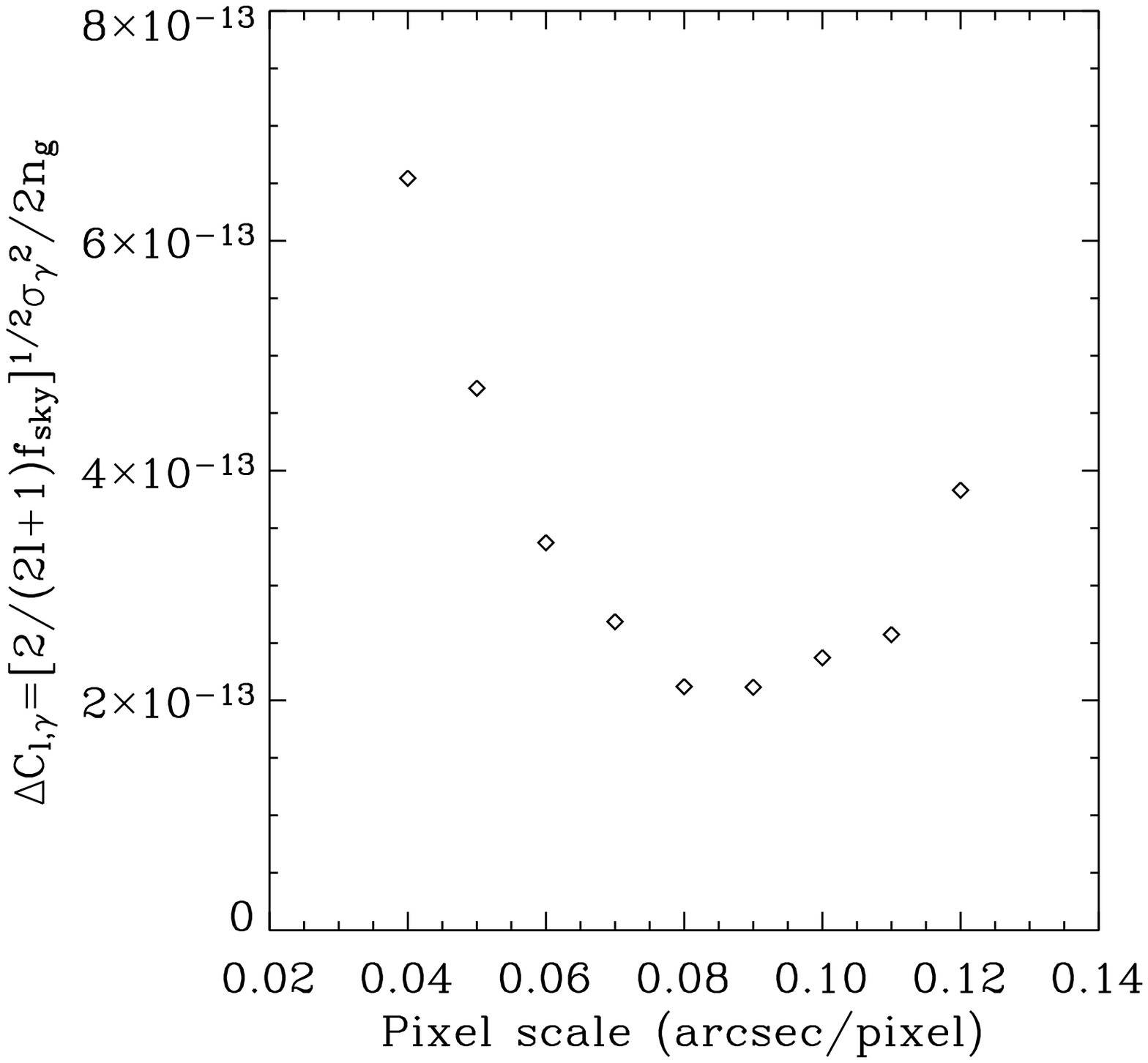} \caption{This plot shows the preliminary
 results of a study aimed at optimizing the pixel scale of the SNAP
 telescope for weak lensing.  The y-axis of the plot shows the
 error $\Delta C_l$ on the matter power spectrum,
 for a fixed survey time and number of pixels.  That is, for a
 smaller pixel scale, the total area surveyed is smaller.  It
 shows that the weak lensing efficiency of SNAP is relatively flat
 for pixels scales near 0.1 arcseconds, the nominal SNAP pixel
 scale.  These results are conservative in that they do not account
for dithering, which would shift the optimum to the right.
 } \label{fig:pixel_size}
\end{figure}

The current baseline SNAP weak lensing survey calls
for imaging 1000 square degrees over the course of roughly one
year. With an extended mission this can be expanded by 1000 square
degrees per year of operation.  SNAP's unique focal plane design
enables a step-and-stare strategy in which each of 6 optical and 3
near infrared filters covers the same patch of sky in a sequence
(Albert et al. 2005). Four 300 second exposures would be taken
with each optical filter (for cosmic ray rejection and better
image sampling).  The NIR filters would get twice the exposure
time.  Due to the high quantum efficiency of the LBL CCDs being
developed for SNAP, this exposure time reaches a magnitude limit
of about 26.6 (AB) in each band for extended sources at a S/N of
10.
The broad spectral
coverage and high photometric accuracy will enable photometric
redshifts for each galaxy with an accuracy $\Delta z < 0.045$ out
to magnitude 26.
The pixel scale of the SNAP CCDs (0.1 arcsecond per pixel)
has been studied using both analytic methods and detailed
image simulations which find that it is well suited for
weak lensing studies with the tight SNAP PSF
(Bernstein 2002; High et al. 2004; Figure~\ref{fig:pixel_size}).

\begin{figure}[h]
\plotone{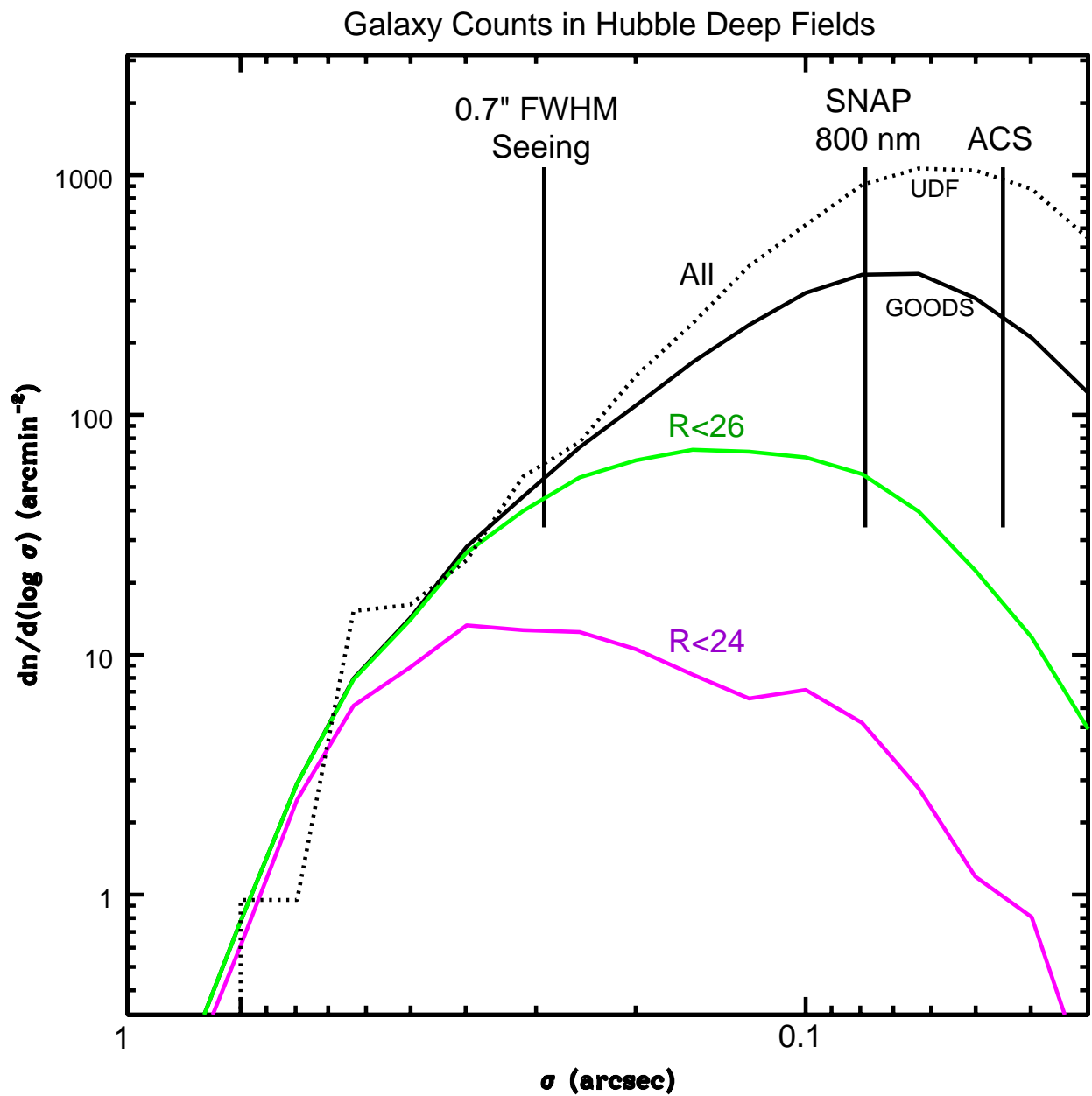} \caption{
The number density of galaxies per square arcminute vs. $\sigma$
as an indicator of galaxy size for a Gaussian profile. These are shown
for different limiting magnitudes as indicated. The vertical lines
show the achievable number density for instruments with
different optical resolution.
 } \label{fig:number}
\end{figure}

A space-based observatory like SNAP will provide significant
improvement in shape measurements.  While the fundamental
limitation of ground-based observing is largely due to atmospheric
blurring (seeing) and sky noise, these do not enter for
space-based measurements. Moreover, SNAP will also benefit from a
more stable thermal environment and the absence of gravity
loading, which will produce a stable PSF, and thus enable superior
shape deconvolution (Rhodes et al. 2004). The noise due to
intrinsic galaxy shapes will be minimized by resolving a high
surface density of galaxies due to the small, stable PSF. Based on
observations with the HST, we estimate SNAP will provide a
threefold increase in the number of resolved galaxies compared to
the best ground based surveys (about 100 galaxies per square
arcminute; Massey et al 2004; Bernstein 2005;
Figure~\ref{fig:number}). Equally important, the photometric
stability and nine filter coverage of SNAP will facilitate
accurate photo-z estimation out to $z\sim 3$.

Note that no space-based WL survey is likely to occur before SNAP;
SNAP will represent a three order of magnitude improvement in this
area over the current largest space-based WL survey, the COSMOS
HST 2-square degree survey.


\section{Systematic Errors}\label{systematics}

\begin{figure}
\psfig{file=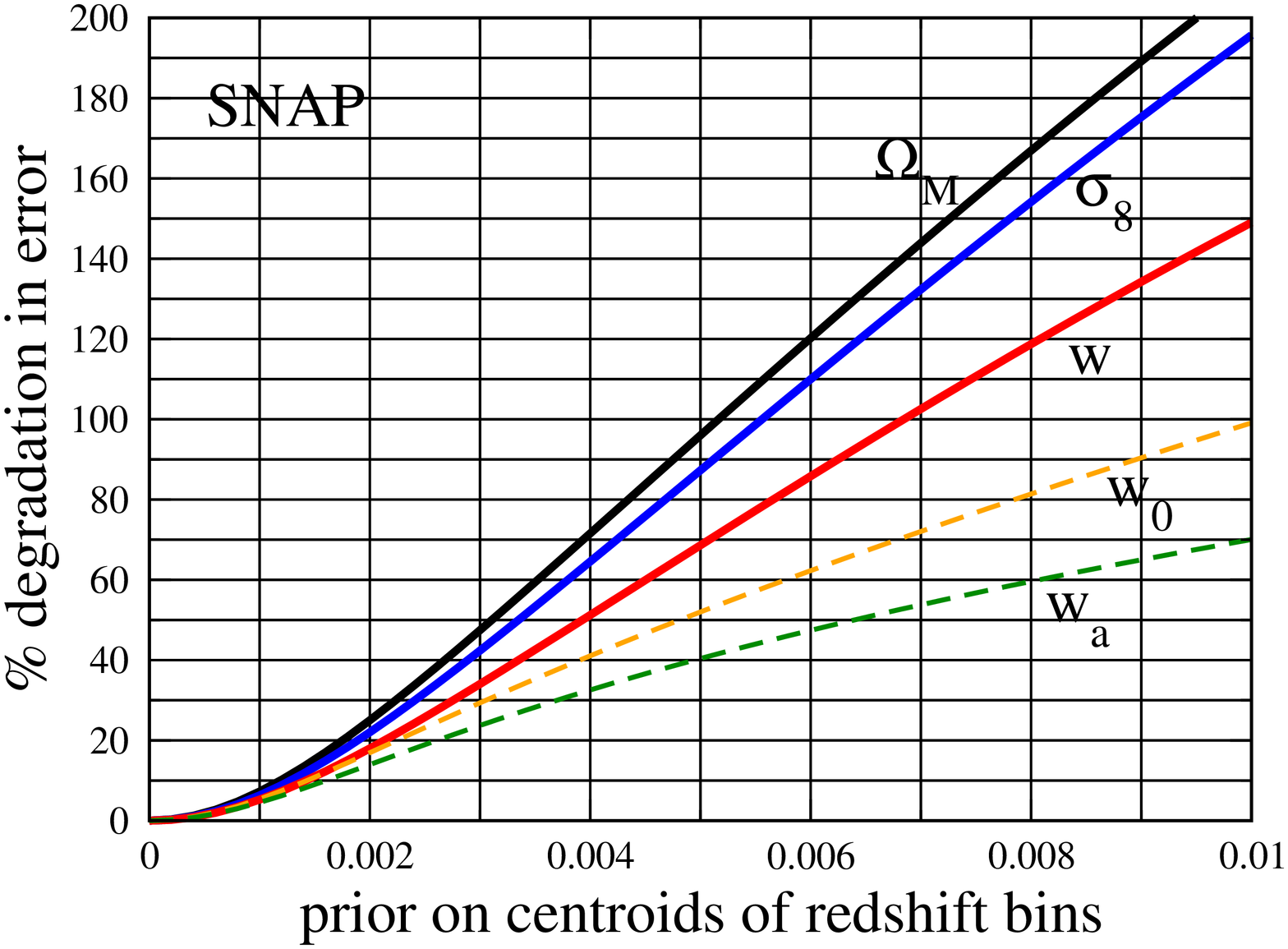, width=3.2in, height=3.5in,
angle=-90}\hspace{-0.2cm} \psfig{file=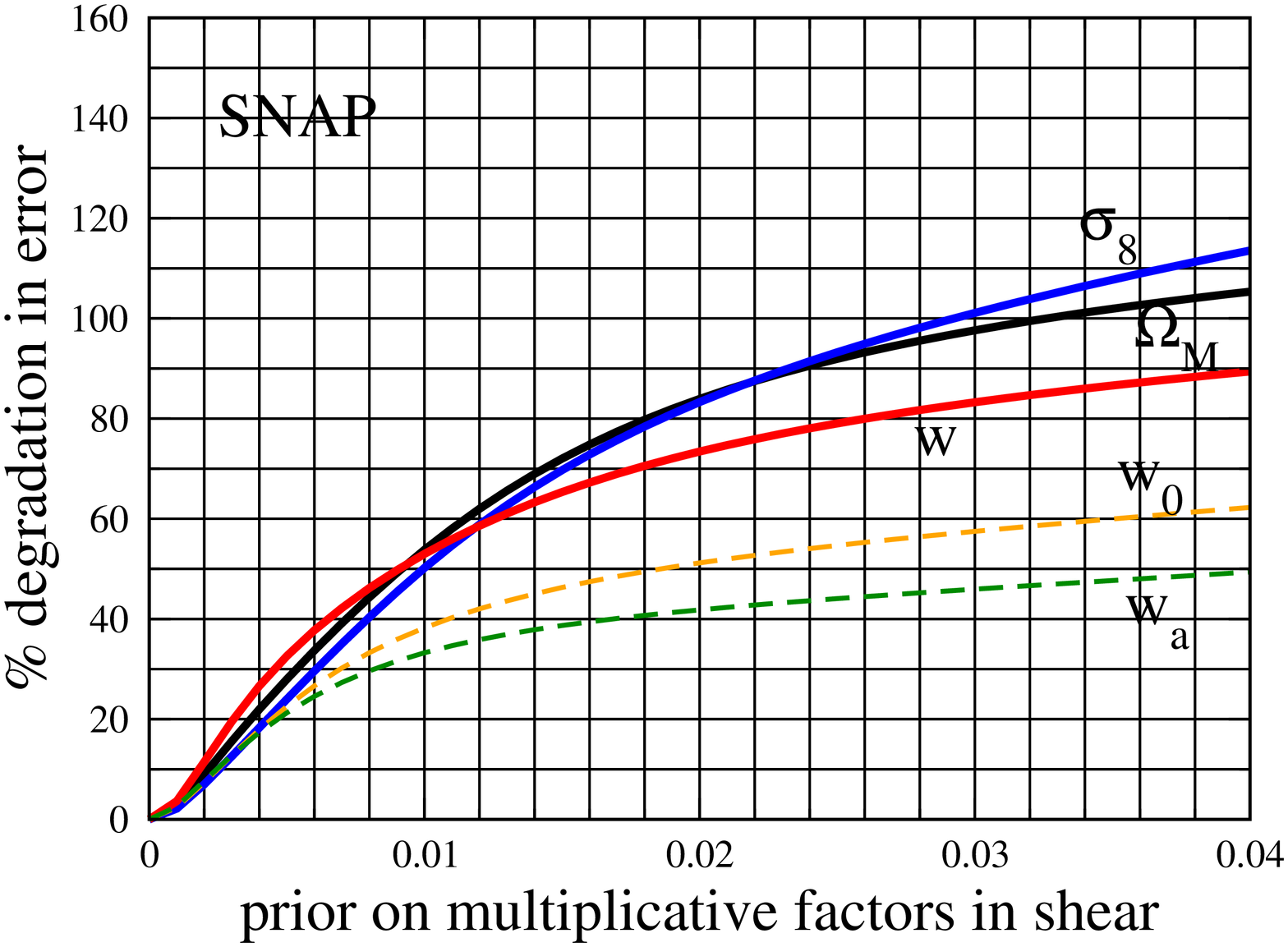,
width=3.2in, height=3.5in, angle=-90} \caption{Requirements on the
redshift biases (left panel) and multiplicative errors (right
panel) for the SNAP weak lensing survey; adopted from Huterer et
al.\ (2005). We show the degradations in (marginalized)
cosmological parameter errors as a function of the redshift and
multiplicative shear bias. Both biases are defined per redshift
bin of $\Delta z=0.3$ and are obtained by averaging individual
biases of galaxies in that bin.  } \label{fig:selfcal}
\end{figure}

Errors in measuring galaxy shapes can roughly be divided into
multiplicative and additive errors (Huterer et al.\ 2005).
The requirement for the SNAP mission is to control multiplicative
error, likely due to errors in shear calibration and PSF
size estimation,
to about $0.01$ or less (or about 1\% of the mean shear in a redshift
bin); see the right panel of Figure \ref{fig:selfcal}.  Additive errors,
which arise due to PSF anisotropy,
require more detailed modeling, but a rough estimate shows that the
contribution needs to be below $0.1\%$ on scales of $\sim 10$
arcmin ($\ell\sim 1000$) in order not to degrade the
error in cosmological parameters.
Furthermore, the redshift distribution of the source galaxies
must be known with a bias $\sim 0.003$ per redshift bin
or smaller. It should be noted that wider area surveys have stricter
requirements on these systematics because the statistical errors are
smaller. For a realistic assessment of the potential of SNAP and ground-based
WL experiments, we must determine how well these systematic errors can
be controlled, and estimate the number of galaxy shapes/redshifts that
can be measured to sufficient accuracy in a given mission.

\subsection{Galaxy Shape Measurement}

\begin{figure}[h]
\plotone{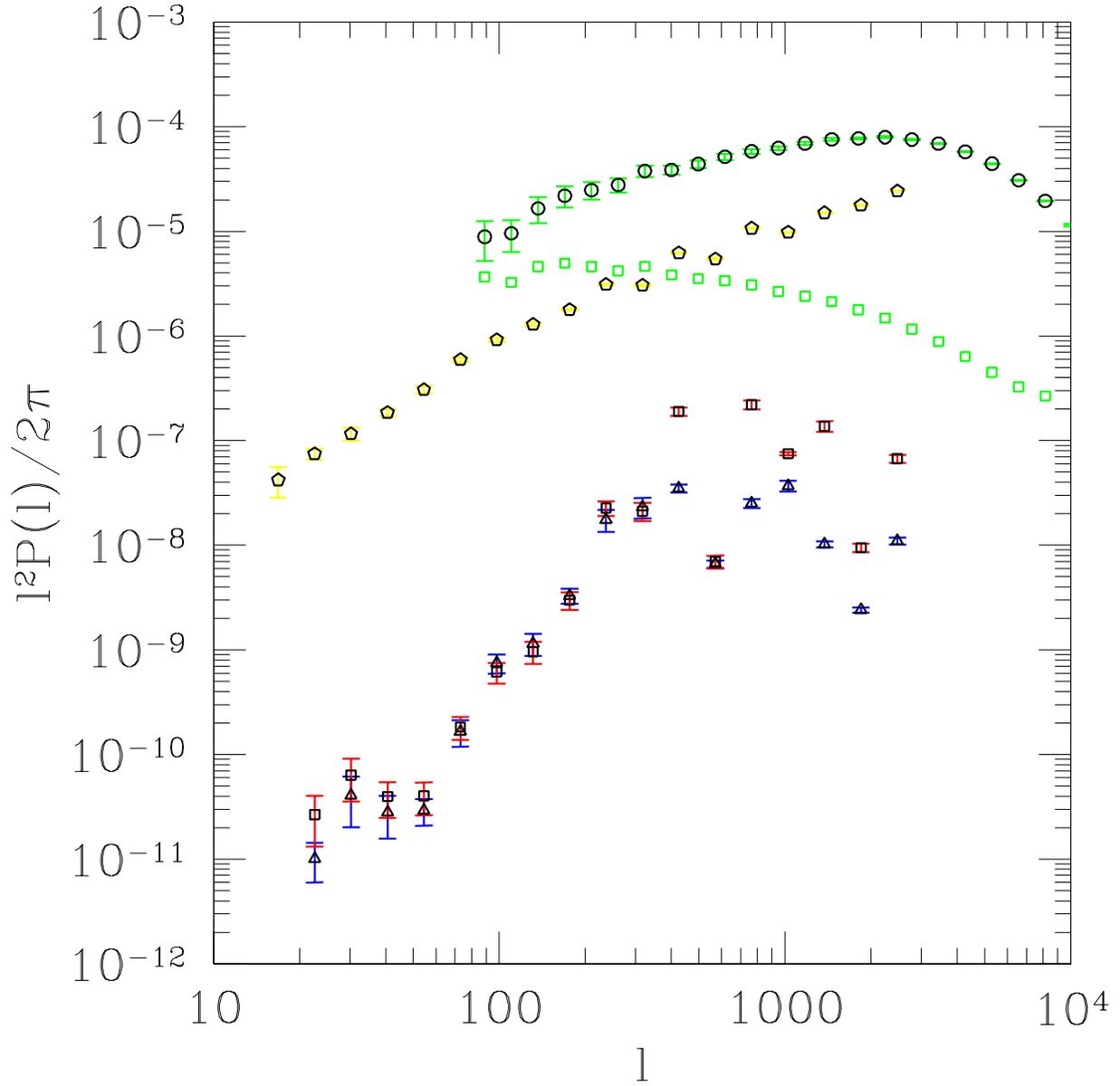} \caption{
 Power spectra of the lensing signal and expected PSF
    anisotropy contamination for the SNAP telescope.  Thermal effects
    and vibration in the structure of the telescope lead to time
    variation in the PSF anisotropy.  This residual power would lead
    to an additive systematic error in the lensing measurement.
    The red curve (squares) is the time-varying power due to thermal
    effects, and has not had the static PSF pattern subtracted off,
    whereas the blue curve (triangles) is the residual power after the
    time invariant pattern is subtracted (as it is expected to be
    measured accurately). The residual power from worst case structural
vibrations is shown in yellow (pentagons).
    The statistical errors for a 1000 square
    degree survey are shown by the green squares, along with the
    power spectrum amplitude of the shear signal from simulations
    (circles).
 } \label{fig:psfpower}
\end{figure}

To estimate the degree of anisotropy of the SNAP PSF, we use a ray
tracer that incorporates all the optical elements of the SNAP
telescope. Misalignments in the optics generate PSF anisotropy,
which we characterize by  ellipticity. The effect of three time
varying effects: thermal drift, guider jitter, and structural
vibration on the PSF have been estimated for expected parameters
of the SNAP telescope.  We generated multiple realizations of a
thousand square degree mock survey  to include the systematic
error pattern induced by these effects. Their contribution to the
power spectrum of the shear is shown in Figure~\ref{fig:psfpower}.
We find that the dominant effects come from guider jitter (which
rises with increasing $\ell$ and thermal drift, which peaks at
$\ell \sim 10^3$). The expected amplitude of the first effect is
comparable to the statistical errors and would require correction.
It is worth noting that ground based telescopes typically have PSF
anisotropy at the percent level, which would lie above the signal
in Figure~\ref{fig:psfpower} and require substantial corrections.

Residual PSF anisotropy can be reduced using measured stellar
PSFs. Hence one must know (a) how many stellar images are
available and how well they allow us to measure the systematics,
and (b) how to correct galaxy images to the highest possible
accuracy. Jarvis \& Jain (2004) have demonstrated a new method
using principal component analysis (PCA) to analyze PSF data that
efficiently recognizes PSF patterns that persist over multiple
exposures.  They show how this led to elimination of detectable
systematic errors in existing ground-based data. We note that the
SNAP PSF requires far smaller corrections than most ground based
instruments; the PCA method is expected to effectively eliminate
the additive PSF errors, since the correction improves with the
number of exposures, which is very large for the SNAP survey.

Answering question (b) requires testing shear calibration.
Shape-measurement algorithms based on the Bernstein \& Jarvis
(2002) and Refregier \& Bacon (2003) approaches are being tested
using simulated data (since no real data can reach this accuracy
yet) to push our accuracy to the sub-percent level that will be
required.

The Shear TEsting Program (STEP) is an ongoing effort by the weak
lensing community to understand the systematic effects and biases
inherent in current methods for measuring galaxy shapes and
deconvolving those shapes from the telescope's PSF.  Heymans et al
(2005) presented the first analysis of the shear measurement
methods of various groups on simulated galaxy images with known
input shears. Subsequent tests on images with more realistic
galaxy morphologies are underway.  These tests will also include
simulations of space-based images.  Of particular importance to
SNAP is the ability of these tests to point out areas in which
current methods can be improved so that the required level of
systematics can be achieved.
\subsection{Photometric Redshifts}

Accurate redshifts of source galaxies are necessary in order to
determine the galaxies' overall distribution in redshift and also
to subdivide them in several redshift bins for lensing tomography
and cross-correlation analyses.  Current photometric redshift
techniques have redshift errors of several percent per galaxy, and
this accuracy is improving as we are learning about the different
techniques. It is estimated that the overall redshift bias
(obtained by averaging over many galaxies in a redshift bin) needs
to be at least an order of magnitude smaller (Huterer et al.\
2005; see left panel of Figure \ref{fig:selfcal}) in order not to
appreciably degrade the error in cosmological parameters. A
similar requirement applies to the scatter in the bias per
redshift bin (Ma et al.\ 2005). Work on understanding and
improving the photometric techniques is ongoing, and the prospects
are excellent as the requisite redshift requirements might already
be achievable even with current techniques.

The broad spectral coverage and high photometric accuracy will
enable photometric redshifts for each galaxy with an rms accuracy
$\Delta z = 0.045$ out to magnitude 26. This is based on estimates
using simulated galaxies and currently well-tested redshift
estimation techniques. We note that if the number of filters is
reduced to six optical filters, then for the same signal-to-noise
the rms photo-z error increases to $0.065$, and has twice the
outlier fraction; and for four optical filters it is $0.1$, with
five times the outlier fraction (the outliers are not used in the
error estimate; Dahlen et al. 2005, in preparation). The expected
levels of bias need further study, as do the requirements of
appropriate spectroscopic samples needed for calibration to reduce
this bias to acceptable levels. This is an important task for
precursor surveys and simulation studies.

\subsection{Theoretical Uncertainties}

For the non-linear power spectrum, we need to know the spectrum in
the range $0.1<k<10\ $Mpc to $1-2\%$ accuracy, with $k\sim 1$
being where the requirement is most stringent (Huterer \& Takada
2005). The current state of the art on N-body simulations is $\sim
5\%$ agreement on the relevant scales for dark matter simulations
with some simulations agreeing better than that.  There is no
known obstacle to reducing the error to the SNAP requirements,
provided only gravitational physics is included. For smaller
scales the effects of baryonic processes are important, though
even if uncorrected they only limit us to $\ell<3000$ (Zhan \&
Knox 2005; White 2005). In the coming decade we can expect to be
able to run thousands of realizations of particle mesh simulations
with $\sim 1$ billion particles and box-lengths of about 500 Mpc,
which would be sufficient for percent-level accuracy in the
lensing power spectrum (Vale \& White 2003).

In order to extend the range of scales which we can use for
constraints on dark energy we will pursue two complementary
strategies.  The first is to improve the modelling. There has been
dramatic progress in modelling the major processes affecting the
matter on sub-Mpc scales, and there is good reason to believe a
small number of simulations including most of the relevant physics
will be available before launch.  Even phenomenological recipes
can help. For example modelling the distribution of gas in
hydrostatic equilibrium in a known potential will go a long way
towards including the effect of hot gas in clusters while
adiabatic contraction models will allow us to model the effect of
the cold component.  Even if these models are not precise, we
should be able to reduce the residual effects by a factor of 2 or
more.

The other route is to use the high quality of the SNAP
observations to mitigate  theory uncertainty.  Since the latter is
dominant on small scales this amounts to using photometric
redshifts of the source galaxies to arrange that the lensing
weight is peaked away from $z=0$ where the small-scale power has
the most influence on any given angular scale.  There exist
several techniques for doing this (Huterer \& White 2005), the
most promising being $k$-space cutting or simply eliminating
source galaxies with low inferred redshifts.

With this knowledge, we would be able to use the small-scale
resolution of SNAP to extract more cosmological information
from the power spectrum and measure new statistics, such as higher
order correlations. In our current forecasts, we have not relied
on small scales that are affected by non-gravitational physics.

\subsection{Tests and Marginalization of Residual Systematics}

The magnitude and scale dependence of B-modes in
the data is a useful monitor of systematics
as they are not produced by scalar gravitational perturbation modes.
There are numerous other tests, including the self-consistency of
the power spectrum and bispectrum, that will be useful to check
for systematic errors.

Weak gravitational lensing offers prospects for
self-calibration (Huterer et al.\ 2005), where weak lensing data is
used to concurrently determine both the systematic and the
cosmological parameters.  The effects of the parameterized
systematics can then be marginalized out without the need to know
their values (but at the expense of increasing the cosmological
parameter errors). A particularly promising approach is to combine
weak lensing methods (for example the PS and BS
measurements) as in those cases the self-calibration may be achieved
with a smaller degradation in cosmological parameter errors.
Further study is needed to model systematics in greater detail
and to allow for shape and redshift errors to be correlated.

\subsection{Weak Lensing Measurements from Space}

\begin{figure}[h]
\plotone{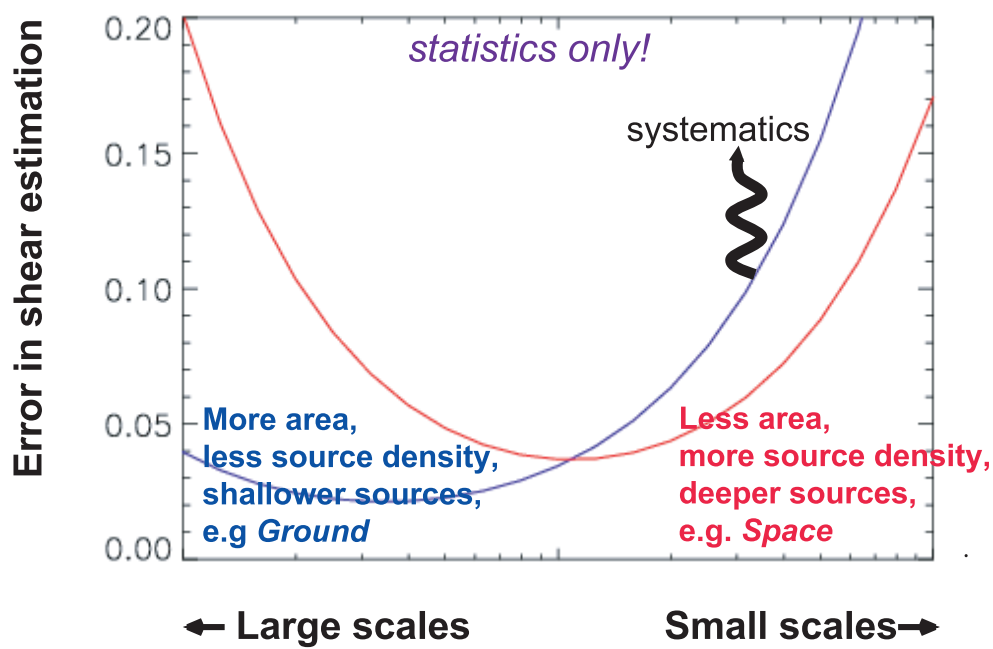} \caption{
 A qualitative representation of the error in shear estimation
 over a range of angular scales from $l=10$ to about $l=10^4$.   At
 the largest angular scales (small $l$) a very wide survey like
 LSST will provide smaller errors.  At higher $l$ the space-based
 survey will have smaller errors due to the higher surface density
 and better shape measurements available.  Systematics, which are
 considerably worse from the ground, will tend to push the blue,
 ground-based curve up.
 } \label{fig:snaplsst}
\end{figure}

One of the crucial advantages of a space-based mission will  be
control of systematic errors that is superior to  that from
ground-based surveys. Since the expected systematic errors from
ground-based telescopes have yet to be precisely quantified, this
advantage will become more evident as work on systematics
progresses. Figure~\ref{fig:snaplsst} shows a representation of
how a space-based survey like SNAP is complementary to a much
larger ground-based weak lensing survey such as LSST.

\begin{figure}[h]
\plotone{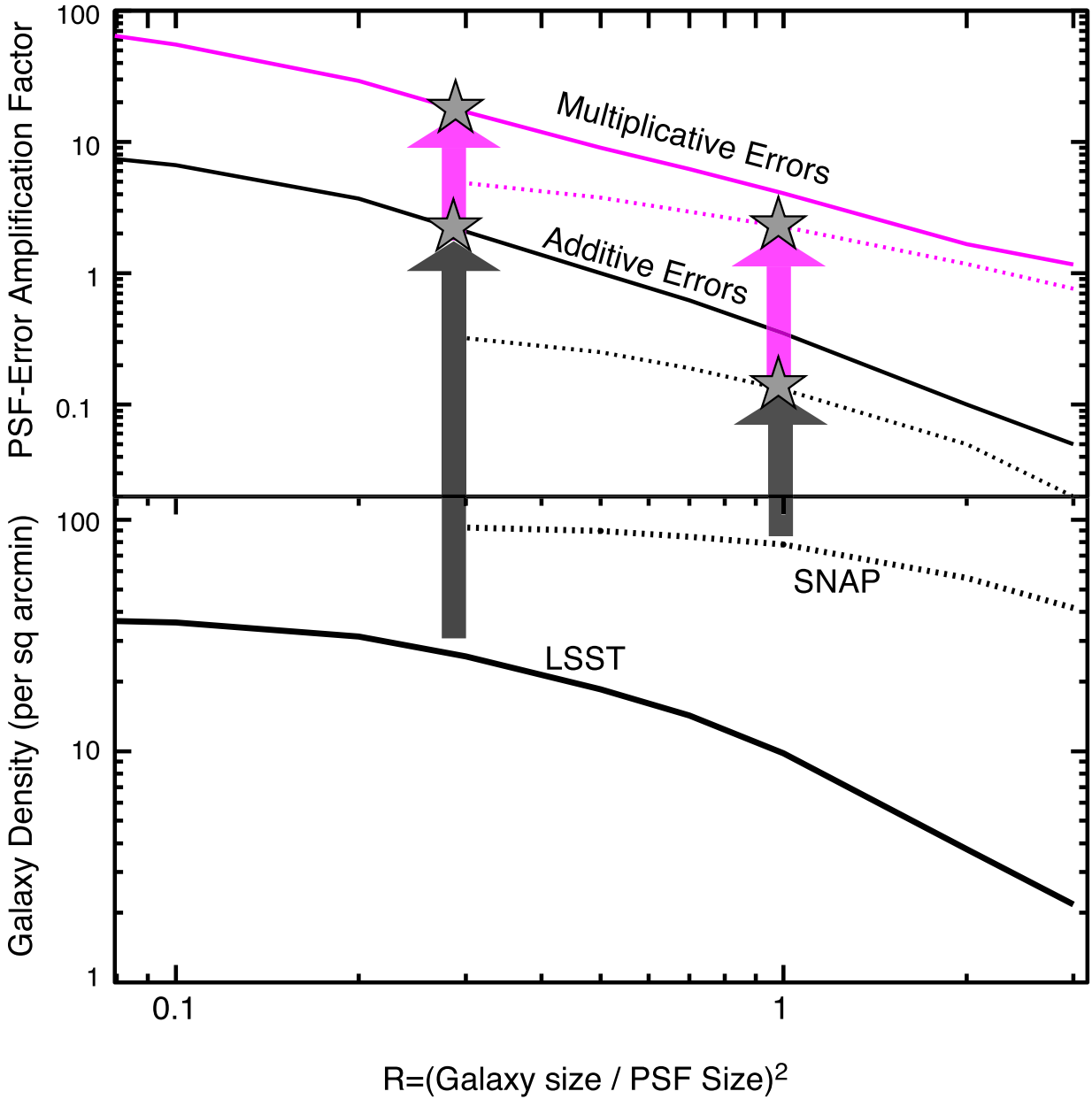} \caption{
Using the Hubble deep fields as a census of the size distribution of
galaxies, the bottom panel shows the effective number of
measurable galaxies vs. the minimum intrinsic galaxy size (relative to
the PSF).  For the
nominal SNAP survey, most galaxies are resolved, so the curve plateaus
below $R=1$.  From the ground, most galaxies are marginally or poorly
resolved, so the curve continues rising to $R<1$.
A plateau is reached
for $R\sim 0.3$ because one requires unattainably high S/N to measure the
shape of more poorly resolved galaxies.  This plateau has ~3x fewer
galaxies per square arcminute than the nominal SNAP Wide survey.
The upper panel plots the factor by which errors in PSF ellipticity
(size) determination are amplified into additive (multiplicative) errors
in the measured lensing shear power spectrum.  As expected, the measured
shear becomes very sensitive to PSF errors when one attempts to use
poorly-resolved galaxies. We see that a space survey, using galaxies
with $R>1$, would have 1-2 orders of magnitude less sensitivity to PSF
errors than a ground survey using $R>0.3$ galaxies.
 } \label{fig:sys}
\end{figure}

Figure~\ref{fig:sys} shows one way to estimate
how the superior imaging of SNAP translates into improved
systematic errors in shear measurements. There is at least
an order of magnitude advantage in the resultant systematics.
The ratio of the typical galaxy size to the PSF is thus critical
not only for attaining a higher number density, but also to
protect against systematic errors.
In addition, a direct source of systematic errors is the anisotropy
of the PSF. The stable thermal environment and
absence of gravity loading will enable SNAP to have a far
better controlled PSF than ground-based telescopes.
Finally, the estimation of photo-z's is greatly aided by the
superior photometric calibration and 9 filters planned for SNAP.
Being able to image with all filters with the same instrument,
and to span the NIR bands, will allow SNAP to attain an rms redshift
error better than $0.05$ out to $z=3$. This is critical, since without
accurate redshift estimates, even accurate shear measurements could
lead to biased estimates of cosmological parameters.

For given sky coverage the statistical errors are significantly
better with SNAP due to the $\sim3$ times higher number density of
well-resolved galaxies. Thus, SNAP imaging will enable high
redshift and small angular scale studies with a far greater
precision than even the best possible ground-based survey. Given
how little we know about dark energy, it is valuable to be able to
probe a wide redshift range to be able to discover new effects in
dark energy evolution.

\section{Relevance of Current and Ongoing Surveys}

Near term, ground based weak lensing surveys will cover hundreds
(Canada-France-Hawaii Telescope Legacy Survey) to thousands
(KiloDegree Survey, RCS2, DES, Pan-Starrs) of square degrees to
moderate depth. These will mature the field of weak lensing for
cosmological constraints.  While the results will lack the
precision and accuracy achievable with SNAP, they will provide
large scale information and the opportunity to test galaxy shear
and photo-z measurement methods. This will lay the groundwork for
the third generation experiment of SNAP, opening the window on the
physics of the universe.

\section{Forecast Dark Energy Constraints from SNAP}

In addition to the strong leverage on cosmological parameter
determination, and robustness in the use of two independent
methods, the combination of WL and supernova distances enables a
fundamental test of the framework: the theory of gravity.  By
combining distance information on the expansion history and WL
information on the growth history, breakdowns or extensions of
Einstein gravity can be distinguished from new physical components
as the physics responsible for acceleration of the universe
(Linder 2005). This is a critical advantage needed for true
understanding of dark energy.

Table 1 shows the constraints on dark energy parameters with a
1000 square degree WL survey combined with the SN measurements.
The parameter analysis is as described in the SN white paper. Note
that significant gains in the WL constraints are expected on
inclusion of bispectrum tomography (Takada \& Jain 2004) and
cross-correlations between foreground galaxies and shear (Jain \&
Taylor 2003; Bernstein \& Jain 2004; Song \& Knox 2004; Hu \& Jain
2004; Zhang et al, 2003). Further, a survey that lasts beyond the
basic 1 year survey would improve the statistical errors which
scale as $f_{\rm sky}$. A 4000 square degree survey (3 additional
years of survey time) would lead to a factor of two improvement in
the constraints from WL as well as being valuable for studies of
baryon oscillations, galaxy clusters, and diverse astronomy.

\begin{deluxetable}{cccccc}
\tablewidth{0pt}
\tablecolumns{5}
\tablecaption{SNAP 1-$\sigma$ uncertainties in dark-energy parameters,
  with conservative systematics for the supernova and a 1000 sq.\
  deg.\ weak-lensing survey. The WL survey does not include
  constraints from the BS or cross correlation.
\label{sci_errors.tab}} \tablehead{Model & $\sigma_{\Omega_w}$ &
$\sigma_{w_0}$ & $\sigma_{w'}$

}
\startdata
Fiducial Universe: flat, $\Omega_M=0.3$, Cosmological Constant dark energy \\
\cline{1-1}\\

SNAP SN + WL; flat; $w(z)=w_0+2w'(1-a)$
                       &  $0.005$   & $0.05$&$0.11$\\
 \\
Fiducial Universe: flat, $\Omega_M=0.3$, SUGRA-inspired dark energy\\
\cline{1-1}\\

SNAP SN + WL; flat; $w(z)=w_0+2w'(1-a)$
                       &  $0.005$  & $0.03$&$0.06$\\
\enddata
\tablecomments{Cosmological and dark-energy parameter
precisions for two fiducial
flat universes with $\Omega_M=0.3$: one in which the dark energy is
attributed to a Cosmological Constant and the other to a
SUGRA-inspired dark-energy model.  The parameter precisions then
depend on the choice of data set, priors from other experiments,
assumptions on the flatness of the universe,
and the model for the behavior of $w$.
In this paper we define $w'\equiv -dw/d\ln a|_{z=1}$}
\end{deluxetable}

\section{References}

\noindent Albert et al. 2005 PASP submitted, astro-ph/0405232

\noindent Bernstein 2005, in preparation

\noindent Bernstein \& Jain, 2004, ApJ, 600, 17

\noindent Bernstein \& Jarvis 2002, ApJ, 583, 123

\noindent Bernstein 2002, PASP, 114, 98

\noindent Dahlen et al, 2005, in preparation


\noindent Hamana et al, 2004, MNRAS 350, 893

\noindent Heymans et al, 2005, MNRAS submitted, astro-ph/0506112

\noindent High et al, 2004, BAAS, 205, 6502

\noindent Hoekstra, 2004, MNRAS, 347, 1337

\noindent Hu \& Jain, PhRvD, 70, 3009

\noindent Huterer et al, 2005, MNRAS submitted astro-ph/0506030

\noindent Huterer \& Takada, 2005, Astroparticle Physics, 23, 369

\noindent Huterer \& White 2005, PRD, submitted, astro-ph/0501451

\noindent Jain \& Taylor 2003, PRL, 91, 141302

\noindent Jarvis \& Jain, 2004, ApJ submitted, astro-ph/0412234

\noindent Linder, 2005 astro-ph/0507263

\noindent Ma, Hu \& Huterer, 2005,  astro-ph/0506614


\noindent Massey et al, 2004, AJ, 127, 3089




\noindent Refreiger \& Bacon, 2003, MNRAS, 338, 48

\noindent Refregier, 2003, ARA\&A, 41, 645

\noindent Rhodes et al, 2004, APh, 20, 377

\noindent Song \& Knox, 2004, PhRvD, 70, 3510

\noindent Takada \& Jain, 2004, MNRAS, 348, 897

\noindent Vale \& White, 2003, ApJ, 592, 699

\noindent White, 2005, APh, 23, 349

\noindent Zhan \& Knox, 2005, ApJ, 616, 75

\noindent Zhang et al, 2003, ApJ submitted, astro-ph/0312348

\newpage

\end{document}